# Multiple carriers of Q noble gases in primitive meteorites

Yves Marrocchi[1,2,*], Guillaume Avice[1,2] & Nicolas Estrade[1,2,3]

[1] Université de Lorraine, Vandoeuvre-lès-Nancy, F-54501, France

[2] CNRS, CRPG, UMR 7358, Vandoeuvre-lès-Nancy, F-54501, France

[3] PCIGR, EOS, University of British Columbia, Vancouver, BC, Canada

[*] Corresponding author's email: yvesm@crpg.cnrs-nancy.fr

Key points

- Noble gas isotopic fractionation
- Origin of meteoritic noble gases
- Evolution of the accretion disk

**Abstract**

The main carrier of primordial heavy noble gases in chondrites is thought to be an organic phase, known as phase Q, whose precise characterization has resisted decades of investigation. Indirect techniques have revealed that phase Q might be composed of two subphases, one of them associated with sulfide. Here we provide experimental evidence that noble gases trapped within meteoritic sulfides present chemically- and thermally-driven behavior patterns that are similar to Q-gases. We therefore suggest that phase Q is likely composed of two subcomponents: carbonaceous phases and sulfides. *In situ* decay of iodine at concentrations levels consistent with those reported for meteoritic sulfides can reproduce the $^{129}$Xe excess observed for Q-gases relative to fractionated Solar Wind. We suggest that the Q-bearing sulfides formed at high temperature and could have recorded the conditions that prevailed in the chondrule-forming region(s).

# 1. Introduction

Primordial noble gases trapped in chondrites are concentrated in residues obtained after demineralization by HF/HCl of the respective bulk meteorites [*Lewis et al.*, 1975]. Most of the heavy noble gases (Ar, Kr & Xe) and a small amount of He and Ne are readily released from the original HF/HCl residues by $HNO_3$ oxidation. This discovery led to the operational definition of phase Q, the oxidisable carrier of primordial noble gases (hereafter Q-gases), which has been found to be ubiquitous in different classes of chondrites [*Busemann et al.*, 2000; *Huss et al.*, 1996]. The nature of phase Q is still under debate but it likely corresponds to carbonaceous structures as noble gas abundances released from acid residues by stepped combustion correlate with those of carbon [*Ott et al.*, 1981]. However, despite the consensus on the carbonaceous nature of phase Q the phase has not yet been isolated from acid residues [*Amari et al.*, 2013]. Nevertheless, indirect techniques have enabled characterization of Q-gases and have revealed: (i) a high noble gas concentration [*Huss et al.*, 1996], (ii) a significant fractionation relative to the solar composition in favor of heavy elements and isotopes [*Busemann et al.*, 2000] and (iii) a common high gas-release temperature for all noble gases in the range 1000-1200°C for unaltered chondrites [*Huss et al.*, 1996]. In addition, several studies indicated that phase Q may consist of two subcomponents: $Q_1$ which is readily soluble in $HNO_3$ and contains most of the heavy noble gases; and $Q_2$ which dissolves slowly in hot concentrated $HNO_3$ [*Busemann et al.*, 2000; *Gros and Anders*, 1977; *Marrocchi et al.*, 2005a]. It has been proposed that at least one of these subcomponents might be related to sulfides [*Gros and Anders*, 1977] but no study has specifically investigated this possibility. However, recent studies report striking results that also suggest that sulfides may have been underestimated as a potential subcarrier of Q-gases. Troilite (FeS) from iron meteorites reproduces the thermal behavior of phase Q well, with a common release temperature of 1000-1200°C for all noble gases [*Nishimura et al.*, 2008]. In addition, stepped combustion measurements on CR chondrites have revealed that very little carbon is associated with Q-gases, suggesting that phase Q might not be solely carbonaceous [*Verchovsky et al.*, 2012]. Furthermore, study of the microdistribution of noble gases within ordinary chondrites has revealed that the sulfide coatings surrounding chondrules exhibit Ne and Ar concentrations at the Q-level as well as Q-like elemental ($^{36}Ar/^{20}Ne$) and isotopic ($^{38}Ar/^{36}Ar$) ratios [*Vogel et al.*, 2004]. Here we report results from an experimental study in which the same chemical treatments as those used for the isolation of phase Q (HF/HCl treatment) and the release of Q-gases ($HNO_3$ oxidation) were applied to iron sulfides separated from the Mundrabilla iron meteorite (IAB). We also test the possibility that sulfides could contribute significantly to the $^{129}Xe$ and $^{131,132,134,136}Xe$ excesses observed for Xe-Q relative to fractionated solar wind.

## 2. EXPERIMENTAL DETAILS

2.6 g of pyrrhotite ($Fe_{0.98}S$) extracted from a fragment of the Mundrabilla iron meteorite (Naturmuseum Senckenberg - Frankfurt - Germany) was ground and separated into two fractions. The first fraction (FeS) was set aside (without undergoing chemical treatment) in order to determine the noble gas content of the Mundrabilla's pyrrhotites. The second fraction was immersed for 24 hours in a HF/HCl, 0.1/1, v/v mixture at 70°C and under nitrogen flow [*Piani et al.*, 2012]. After HF/HCl treatment, the sample was washed with water, thoroughly dried at 50°C and then separated into two equal aliquots. One aliquot was retained for noble gas analysis (hereafter FeS-HCl) while the other was etched with 14 M $HNO_3$ for 24 hours at 70°C and under nitrogen flow [*Lewis et al.*, 1975]. The resulting etched residue (FeS-$HNO_3$) was washed with water and dried at 50°C before analysis. The FeS-$HNO_3$ fraction exhibits a reddish color that is distinct from the gray color of both the starting material and the HF/HCl residue.

Ar, Kr and Xe were measured in the raw sample, HF/HCl residue, and nitric-etched residue. Samples were weighed, wrapped in platinum foil and then loaded into a glass sample-tree. The samples were gently baked at 150°C for three days in order to remove adsorbed atmospheric gases. Noble gases were extracted by stepped pyrolysis in the temperature range 300-1768°C using a tungsten coil. The linear current-temperature calibration curve for the coil was obtained using an optical pyrometer with a precision of ± 25°C for temperatures above 800°C. The calibration curve was extrapolated to temperatures lower than 800°C. Extraction times were adjusted as function of temperature: (i) 15 min for the four low-temperature steps (i.e., 300-1300°C), (ii) 12 min for the 1500°C step and (iii) 5 min for the final step (at the platinum melting-point: 1768°C). The released gases were exposed to three consecutive pellet getters containing SAES St172 getter alloy in order to remove active gases (10 min at 450°C, 10 min at room temperature). Ar, Kr and Xe were held on a charcoal finger at liquid nitrogen temperature for 45 minutes, then the residual light noble gases were pumped out over 5 min. Active charcoal immersed in liquid nitrogen was then heated to -105°C using an electric wire that surrounds the charcoal. Calibration curves determined from air standards showed that Kr and Xe are not affected at this temperature while 70 % of the Ar is released. This procedure reduces the amount of Ar in the mass spectrometer, which has a considerable effect on the mass discrimination of krypton and xenon. Argon released from the charcoal finger was then pumped out for 5 min. The liquid nitrogen was then removed and the charcoal finger was heated to 250°C for 25 min to release Xe, Kr and the residual fraction of Ar. Heavy noble gases were then introduced into high sensitivity pulse-counting static mass spectrometer (Washington University, Saint Louis, USA, [*Mabry et al.*, 2007; *Meshik et al.*, 2007]) for determination of Ar, Kr, and Xe abundances and $^{38}Ar/^{36}Ar$, $^{86}Kr/^{84}Kr$, and $^{129}Xe/^{132}Xe$ isotopic ratios.

Hot blanks (1200°C) were performed several times during each analytical session. The Kr and Xe concentrations within Pt foils were also measured and appeared to be negligible. The measured Kr and Xe abundances are typically accurate to better than 5% whereas the Ar concentrations present a lower precision of ≈ 25% and are not presented here. The uncertainties in the $^{86}$Kr/$^{84}$Kr and $^{129}$Xe/$^{132}$Xe isotopic ratios (1σ) include hot blank, standard and sample uncertainties (Table 1).

## 3. RESULTS

The stepwise-heating analysis show that pyrrhotite releases Kr and Xe at temperatures between 900 and 1768°C, with a maximum release occurring in the range 900-1175°C (Fig 1; Table 1). The total concentration of $^{132}$Xe measured in the Mundrabilla's pyrrhotite was 3.27 $10^{-10}$ cm$^3$ STP.g$^{-1}$, while the $^{84}$Kr concentration was slightly higher (9.21 $10^{-10}$ cm$^3$ STP.g$^{-1}$). $^{86}$Kr/$^{84}$Kr ratios measured for each temperature step present homogenous values close to the atmospheric composition (Table 1). $^{129}$Xe/$^{132}$Xe ratios show more important variations with a clear excess of radiogenic $^{129}$Xe in all except the final extraction steps that lies close to the atmospheric composition (Table 1). These results are consistent with previous reports of Kr and Xe concentrations within troilites from iron meteorites [*Mathew and Marti*, 2009; *Nishimura et al.*, 2008]. No significant mass variation was observed following the HF/HCl treatment but the FeS-HCl residues displayed a significant gas loss compared to the initial material, both for Kr (≈53%) and Xe (≈38%; Fig 1; Table 1). However, Kr and Xe did present the same thermal release patterns as the original sample, with the maximum noble gas release occurring in the range 1000-1200°C (Fig. 1). $^{86}$Kr/$^{84}$Kr and $^{129}$Xe/$^{132}$Xe ratios are similar to the original sample (Table 1). The FeS-HNO$_3$ residue displayed a reddish color that was distinct from the gray color of the starting material and the HF/HCl residue, but no mass variation was observed compared to the initial material. The amounts of Kr and Xe released from the FeS-HNO$_3$ residue are much lower than those from the FeS-HCl fraction, with gas losses of 65% and 69% recorded for Kr and Xe, respectively (Fig 1; Table 1). Stepwise-heating analyses revealed that all of the temperature steps were affected by the degassing of Kr and Xe (Fig. 1). $^{129}$Xe/$^{132}$Xe ratios show resolvable but lower excess of radiogenic $^{129}$Xe comparing to the other residues (Table 1).

## 4. DISCUSSION

Significant degassing was observed upon HF/HCl treatment, suggesting that sulfides are sensitive to chemical digestion (Fig. 1). However, HF/HCl-treated pyrrhotite still contains a significant amount of Kr and Xe and could thus be considered as an acid-resistant mineral. This is confirmed by TEM observations of meteoritic acid residues that contain an important amount of inorganic material such as sulfides and oxides [*Derenne and Robert*, 2010]. In addition, our results indicate that pyrrhotite has a strong susceptibility to HNO$_3$ oxidation (Table 1). This feature has also been reported for pentlandite (i.e., (Fe,Ni)$_9$S$_8$) but pentlandite differs from phase Q in its thermal stability with decomposition starting at 610°C [*Kerridge et al.*, 1979]. In contrast, the release pattern of noble gases trapped within pyrrhotite closely resembles that of the Q-gases, with a major release of Kr and Xe occurring between 900 and 1200°C (Fig 1). Similar results were obtained for each of the five noble gases released from troilite from the Saint Aubin iron meteorite (UNGR) [*Nishimura et al.*, 2008], demonstrating that nickel-poor iron sulfides match the thermal characteristics of phase Q. The high temperature release is directly linked to the incongruent dissociation of pyrrhotite, which occurs in the range 1000-1200°C, depending on the stoechiometry [*Kellerud*, 1963]. Hence, our results suggest that chondritic sulfides could (i) represent a plausible subcarrier of Q-gases and could (ii) contribute to the Q-gas budget at the maximum-release temperature of Q-gases (i.e., 1000-1200°C) [*Huss et al.*, 1996]. This is in good agreement with sulfides separated from Allende (CV3) that show typical Xe-Q isotopic compositions at high temperature (i.e., > 900°C), representing few percent of the total Xe-Q reported for this chondrite [*Busemann et al.*, 2000; *Lewis et al.*, 1977]. It might be argued that the amounts of Kr and Xe measured in iron sulfides from the Mundrabilla iron meteorite are three orders of magnitude lower than those reported in phase Q [*Busemann et al.*, 2000; *Huss et al.*, 1996]. However, the goal of this study was not to isolate phase Q as iron meteorites do not show this noble gas component (except in rare graphite nodules, [*Matsuda et al.*, 2005]) but rather to test the chemical sensitivity and the thermal behavior of noble gas-bearing sulfides. Consequently, we suggest that phase Q likely corresponds to complex Q-gas subcarriers of different natures: carbonaceous phases and iron sulfides.

Among the meteoritic noble gases, Xe-Q in different classes of chondrite is characterized by mass-dependent fractionation relative to solar wind (SW), favoring the heavy isotopes [*Marrocchi and Marty*, 2013; *Meshik et al.*, 2014]. However, clear excesses of $^{129}$Xe and $^{131,132,134,136}$Xe are also observed together with the mass-fractionated SW (Fig. 2a). These features are generally explained by a mixing model in which 98.4% of ≈ 8 ‰.amu$^{-1}$ mass-fractionated SW-Xe is mixed with 1.6% Xe-HL from nanodiamonds and monoisotopic $^{129}$Xe from $^{129}$I decay [*Gilmour*, 2010]. However, Xe-HL cannot contribute to Xe-Q release generated by on-line nitric oxidation of acid residues [*Busemann et al.*, 2000] because nanodiamonds are unaffected by etching [*Crowther and Gilmour*, 2013]. Hence, the $^{131,132,34,136}$Xe excesses have been attributed to a gas carrier that presents the same enrichment in the light and heavy xenon isotopes as nanodiamonds but which has not yet been isolated [*Gilmour*, 2010]. Given our results, we can test an alternative explanation based on *in situ* fission and decay of $^{238}$U+$^{244}$Pu and $^{129}$I within sulfides. The mass fractionation of Xe-Q relative to SW-Xe and its associated uncertainty were determined for eight chondrites and for the average Q composition [*Busemann et al.*, 2000] using only the non-radiogenic/fissiogenic

Xe-Q isotopic ratios (i.e., $^{124,126,128}$Xe/$^{130}$Xe). According to the abundance of $^{129}$Xe in phase Q [*Busemann et al.*, 2000], and assuming an initial solar system ratio of 1.1 10$^{-4}$ for $^{129}$I/$^{127}$I [*Gilmour et al.*, 2006], we thus determined the iodine content required to correct the $^{129}$Xe excess relative to the fractionated SW (Table 2). Then, on the basis of the abundances of $^{131,132,134,136}$Xe in phase Q [*Busemann et al.*, 2000], we calculated the $^{238}$U concentrations required to correct the excesses of $^{131,132,134,136}$Xe relative to fractionated SW using the fission yields reported for $^{238}$U and $^{244}$Pu [*Ragettli et al.*, 1994], the branching ratio for $^{238}$U and $^{244}$Pu [*Ozima and Podosek*, 2002], the initial solar system ratios of ($^{244}$Pu/$^{238}$U)$_0$ = 6.8 10$^{-3}$ and ($^{238}$U/$^{235}$U)$_0$ = 137.88 [*Ozima and Podosek*, 2002] and a start of radioactive decay 4.57 Gyr ago. Our results show that the respective $^{129}$I and $^{238}$U initial concentrations required to correct the $^{129}$Xe and $^{131,132,134,136}$Xe excesses relative to fractionated SW observed for Xe-Q across different chondrites [*Busemann et al.*, 2000] fall in the range of 0.03-0.45 ppm $^{129}$I and 0-50 ppm $^{238}$U (Table 2). The resulting iodine contents are in good agreement with the concentrations of 0.1-3.5 ppm reported for sulfides from different types of meteorites [*Clark et al.*, 1967; *Goles and Anders*, 1962]. Consequently, we propose that Q-gases trapped within sulfides could be responsible for the $^{129}$Xe excess observed during the release of Q-gases. In contrast, the uranium contents required within sulfides to explain the $^{131,132,134,136}$Xe excess relative to fractionated SW are generally too high compared to the sub-ppm concentrations reported for meteoritic sulfides [*Crozaz*, 1979]. Moreover, the Xe-Q corrected for $^{238}$U+$^{244}$Pu contributions does not produce a better fit for heavy xenon isotopes than the canonical model involving mixing of different Xe reservoirs [*Meshik et al.*, 2014] (Fig. 2b). Consequently, the excesses of $^{129}$Xe could be attributed to the *in-situ* decay of $^{129}$I within sulfides, while fission of $^{238}$U+$^{244}$Pu would play a negligible role in generating the excess of $^{131,132,134,136}$Xe relative to fractionated SW observed for Q-gases (Fig. 2a).

Q-gases are ubiquitous among the different types of chondrites, despite the fact that they experienced diverse secondary alteration processes such as fluid percolation and/or metamorphism [*Bourot-Denise et al.*, 2010; *Brearley*, 2006; *Hewins et al.*, 2014; *Huss et al.*, 2006; *Marrocchi et al.*, 2014]. This suggests that the formation of the S-rich subcarrier of Q-gases is linked to primary high-temperature processes rather than to parent body alteration. Recent reports have revealed that chondrule formation took place under high partial pressure of sulfur, leading to the formation of iron sulfide from the chondrule melts by solubility/saturation processes [*Marrocchi and Libourel*, 2013] and/or condensation at the surface of chondrules [*Tachibana and Huss*, 2005]. The formation of Q-bearing sulfides would be directly related to the chondrule-forming event. Such a view supports models which postulate that the formation of chondrules took place in an environment characterized by volatile-enriched gas that interacts at high temperature with chondrule precursor [*Marrocchi and Libourel*, 2013]. Thus, formation of Q-bearing sulfides can be achieved in regions characterized by enhancement of the respective noble gas partial pressures and under ionizing conditions that allow the isotopic fractionation observed for Q-gases relative to SW to be reproduced [*Hohenberg et al.*, 2002; *Marrocchi et al.*, 2005b; *Marrocchi et al.*, 2011].

## 5- Concluding remarks

We have performed an experiment to test whether iron sulfides might represent a plausible subcomponent of the main noble gas carriers in primitive meteorites - phase Q. Although significant noble gas degassing was observed upon HF-HCL treatment, our results show that noble gases trapped within sulfides present similar chemical susceptibility and thermal behavior than Q-gases. Hence, we propose that sulfides likely represent a plausible subcomponent of phase Q. Under this hypothesis, phase Q represents a mix of multiple primordial noble gas carriers of different natures such as carbonaceous phases and iron sulfide minerals. This suggests that Q-gases may represent a ubiquitous noble gas reservoir outside the Sun at the time of the formation and accretion of the first solids in the protosolar nebula.

**Acknowledgments**

We are grateful to Maïa Kuga, Laurette Piani, Pete Burnard, Pierre-Henri Blard, Bernard Marty, Alice Williams, Barbara Marie, Laurent Rémusat and Matthias M.M. Meier for helpful discussions. The data for this paper are available by contacting Yves Marrocchi (yvesm@crpg.cnrs-nancy.fr). This is CRPG contribution #2333.

| Samples | Mass (g) | Temperature (°C) | $^{84}$Kr ($10^{-10}$ cc.g$^{-1}$) | $^{86}$Kr/$^{84}$Kr | $^{132}$Xe ($10^{-10}$ cc.g$^{-1}$) | $^{129}$Xe/$^{132}$Xe |
|---|---|---|---|---|---|---|
| **FeS** | 0.0149 | 295 | bdl | - | bdl | - |
| | | 591 | bdl | - | bdl | - |
| | | 887 | 1.37 | 30.6 ± 0.4 | 0.76 | 194.6 ± 0.5 |
| | | 1183 | 4.32 | 29.8 ± 0.3 | 1.82 | 245.5 ± 0.4 |
| | | 1478 | 3.52 | 28.9 ± 0.3 | 0.68 | 173.3 ± 0.5 |
| | | 1770 | 1.18 | 30.5 ± 0.4 | 0.21 | 103.6 ± 0.8 |
| | | Total | 9.21 | 29.6 ± 0.3 | 3.27 | 218.5 ± 0.4 |
| **FeS-HCl** | 0.0173 | 291 | bdl | - | bdl | - |
| | | 583 | bdl | - | bdl | - |
| | | 875 | 1.18 | 30.2 ± 0.4 | 0.78 | 164.6 ± 0.6 |
| | | 1166 | 2.07 | 29.9 ± 0.4 | 1.06 | 241.4 ± 0.4 |
| | | 1458 | 1.08 | 29.8 ± 0.4 | 0.17 | 134.4 ± 0.8 |
| | | 1770 | 0.82 | 30.5 ± 0.5 | 0.009 | 113.1 ± 1.1 |
| | | Total | 4.33 | 30.0 ± 0.4 | 2.01 | 202.4 ± 0.5 |
| **FeS-** | 0.0126 | 291 | bdl | - | bdl | - |

| | | | | |
|---|---|---|---|---|
| **HNO₃** | | | | |
| 583 | bdl | - | bdl | - |
| 875 | 0.61 | 30.9 ± 0.6 | 0.21 | 129.4 ± 1.8 |
| 1166 | 0.48 | 30.4 ± 0.6 | 0.21 | 141.2 ± 1.1 |
| 1458 | 0.43 | 30.0 ± 0.6 | 0.20 | 116.6 ± 1.2 |
| 1770 | 0.16 | 30.5 ± 0.7 | 0.16 | 106.4 ± 1.3 |
| Total | 1.52 | 30.5 ± 0.6 | 0.62 | 129.2 ± 1.3 |

Table 1: $^{84}$Kr and $^{132}$Xe concentrations and $^{86}$Kr/$^{84}$Kr and $^{129}$Xe/$^{132}$Xe isotopic ratios determined by stepwise heating of original pyrrhotite (FeS), HF/HCl-treated pyrrhotite (FeS-HCl) and nitric-etched pyrrhotite (FeS-HNO₃). Isotopic ratios ×100. bdl = below detection limit.

|  | Allende | Chainpur | Bokkeveld | Dimmit | Grosnaja | Isna | Lance | Murchison | Cold Q |
|---|---|---|---|---|---|---|---|---|---|
|  | CV3 | LL3.4 | CM2 | H3.7 | CV3 | CO3.7 | CO3.4 | CM2 |  |
| I (ppm) | 0.445 | 0.057 | 0.294 | 0.031 | 0.152 | 0.402 | 0.035 | 0.395 | 0.21 |
| U (ppm) | 50.07 | 2.06 | 0.02 | 0.56 | 5.53 | 33.98 | 2.55 | 43.29 | 9.63 |
| $^{129}$Xe from I decay (cc/g) | 8.64 (-9) | 1.03 (-9) | 5.71 (-9) | 6.13 (-10) | 2.97 (-9) | 7.80 (-9) | 6.90 (-10) | 7.67 (-9) | 4.08 (-9) |
| $^{131}$Xe from U fission (cc/g) | 6.25 (-12) | 2.57 (-13) | 2.49 (-15) | 7.04 (-14) | 6.90 (-13) | 4.24 (-12) | 3.19 (-13) | 5.40 (-12) | 1.20 (-12) |
| $^{132}$Xe from U fission (cc/g) | 4.89 (-11) | 2.01 (-12) | 1.95 (-14) | 5.51 (-13) | 5.40 (-12) | 3.32 (-11) | 2.49 (-12) | 4.23 (-11) | 9.41 (-12) |
| $^{134}$Xe from U fission (cc/g) | 6.84 (-11) | 2.81 (-12) | 2.73 (-14) | 7.71 (-13) | 7.55 (-12) | 4.64 (-11) | 3.49 (-12) | 5.91 (-11) | 1.32 (-11) |
| $^{136}$Xe from U fission (cc/g) | 8.22 (-11) | 3.38 (-12) | 3.28 (-14) | 9.26 (-13) | 9.08 (-12) | 5.58 (-11) | 4.19 (-12) | 7.11 (-11) | 1.58 (-11) |
| $^{129}$Xe from Pu fission | 1.08 (-10) | 4.43 (-12) | 4.30 (-14) | 1.21 (-12) | 1.19 (-11) | 7.30 (-11) | 5.49 (-12) | 9.30 (-11) | 2.07 (-11) |

| | (cc/g) | | | | | | | | |
|---|---|---|---|---|---|---|---|---|---|
| [131]Xe from Pu fission | | | | | | | | | |
| (cc/g) | 5.56 (-10) | 2.29 (-11) | 2.22 (-13) | 6.27 (-12) | 6.14 (-11) | 3.77 (-10) | 2.84 (-11) | 4.81 (-10) | 1.07 (-10) |
| [132]Xe from Pu fission | | | | | | | | | |
| (cc/g) | 2.00 (-9) | 8.24 (-11) | 8.00 (-13) | 2.26 (-11) | 2.21 (-10) | 1.36 (-9) | 1.02 (-10) | 1.73 (-9) | 3.85 (-10) |
| [134]Xe from Pu fission | | | | | | | | | |
| (cc/g) | 2.08 (-9) | 8.58 (-11) | 8.33 (-13) | 2.35 (-11) | 2.30 (-10) | 1.42 (-9) | 1.06 (-10) | 1.80 (-9) | 4.01 (-10) |
| [136]Xe from Pu fission | | | | | | | | | |
| (cc/g) | 2.24 (-9) | 9.23 (-11) | 8.95 (-13) | 2.53 (-11) | 2.48 (-10) | 1.52 (-9) | 1.14 (-10) | 1.94 (-9) | 4.31 (-10) |

Table 2: Amounts of initial $^{238}$U and $^{129}$I required for correcting observed excesses of $^{129}$Xe and $^{131,132,134,136}$Xe in the Xe-Q of different chondrites and in the average Q composition relative to mass-fractionated SW-Xe (Data from [*Busemann et al.*, 2000; *Wieler et al.*, 1992]). Amounts of $^{129}$Xe produced by the decay of $^{129}$I at the concentrations required for correcting $^{129}$Xe excesses observed relative to fractionated SW calculated from non-radiogenic/non-fissiogenic xenon isotopes. $^{129,131,132,134,136}$Xe concentrations produced by the fission of $^{238}$U and $^{244}$Pu concentrations required to correct the observed excesses of $^{131,132,134,136}$Xe in phase Q relative to mass-fractionated SW calculated from non-radiogenic/non-fissiogenic xenon isotopes and iodine-corrected $^{129}$Xe (i.e., $^{124,126,128,129}$Xe/$^{130}$Xe). The number in bracket corresponds to the power.

**Figure captions**

Fig. 1: (a) $^{84}$Kr elemental abundances determined by stepwise heating in the range 300-1768°C for original pyrrhotite (FeS), pyrrhotite that has undergone HF/HCl treatment (FeS-HCl) and etched pyrrhotite (FeS-HNO$_3$). (b) $^{132}$Xe elemental abundances determined by stepwise heating in the range 300-1768°C for original pyrrhotite (FeS), HF/HCl residue (FeS-HCl) and pyrrhotite oxidized by HNO$_3$ (FeS-HNO$_3$).

Fig. 2: (a) Xe-Q composition normalized to the SW-Xe. Clear excesses of $^{129}$Xe and $^{134,136}$Xe are observed relative to a ≈ 10.1 ‰.amu$^{-1}$ mass-fractionated SW-Xe calculated from non-radiogenic/non-fissiogenic xenon isotopes (red line). These excesses are generally attributed to $^{129}$I decay and to a contribution from an unidentified carrier with the same Xe-HL isotopic composition as HL-bearing nanodiamonds. (b) Modeled composition of Xe-Q corrected for the radiogenic and fissiogenic contributions of 0.21 ppm of $^{129}$I and 9.63 ppm of $^{238}$U, respectively. This model is compared to a recent model based on the mixing of fractionated SW-Xe with Xe-HL and Xe-S [*Meshik et al.*, 2014].

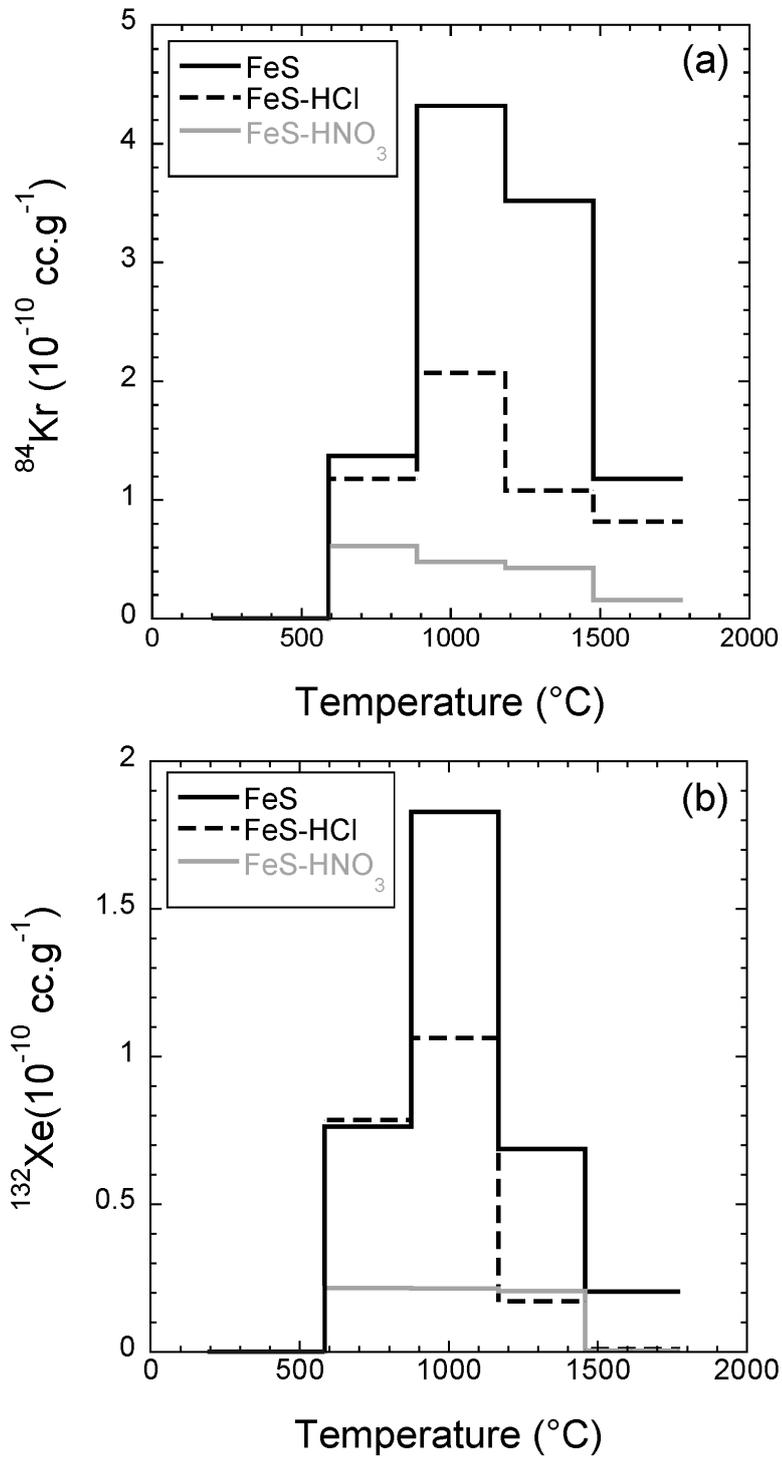

Fig. 1

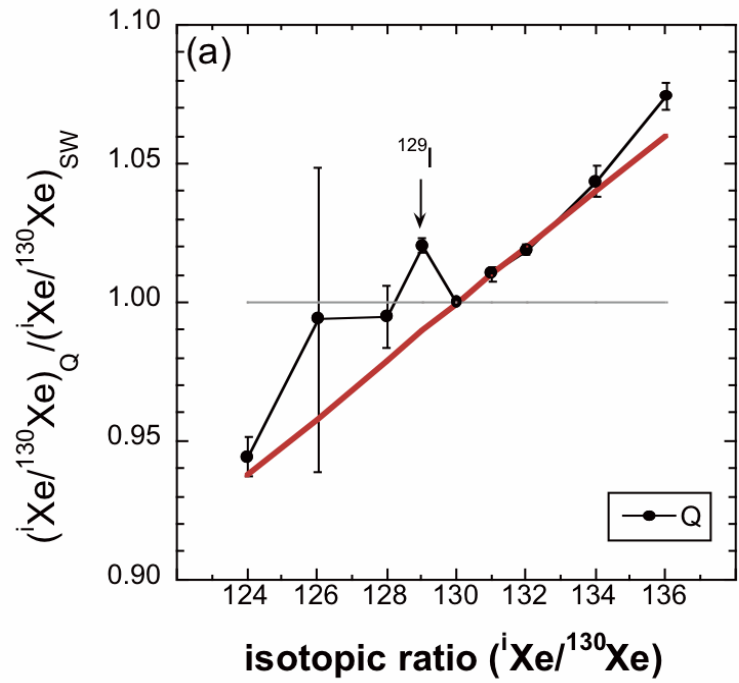

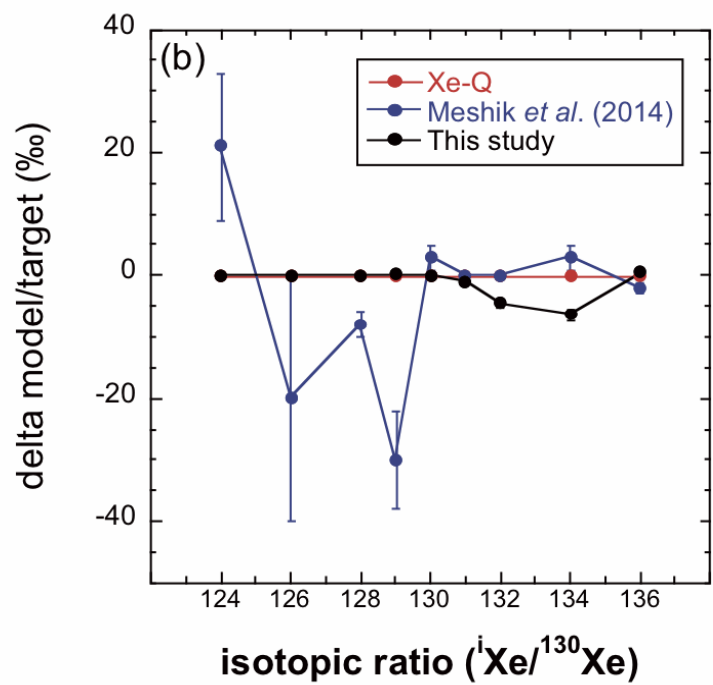

Fig. 2